\documentclass[aps,amsfonts,pra,twocolumn,showpacs]{revtex4}
\usepackage{epsfig,amsmath,amssymb,bm,epsf,graphics,psfrag,verbatim,subfigure}

\def\beq{\begin{equation}}
\def\eeq{\end{equation}}
\def\bea{\begin{eqnarray}}
\def\eea{\end{eqnarray}}

\begin{document}

\title{%
BEC-BCS crossover and the mobility edge: superfluid-insulator transitions and reentrant superfluidity in disordered Fermi gases
}


\author{Sarang Gopalakrishnan}

\affiliation{%
Department of Physics, Harvard University, Cambridge, Massachusetts 02138}

 \date{Dec 2, 2012} 

\begin{abstract}
A superfluid-insulator transition is known to occur in strongly disordered Fermi gases, in both the BCS and BEC regimes; here, we address the properties of this transition across the BEC-BCS crossover. We argue that the critical disorder strength at which superfluidity is lost changes non-monotonically with detuning from Feshbach resonance, and that a reentrant superfluid phase arises for detunings near the fermionic mobility edge. Our analysis of the intermediate regime is quantitatively valid for narrow resonances and near four dimensions, and provides a simple physical picture of this regime, in terms of two distinct but coexisting insulators.
\end{abstract}

\maketitle

\section{Introduction}

The physics of interacting electrons in strongly disordered media involves a delicate interplay between Cooper pairing, Anderson localization, and Coulomb repulsion. The nature of the resulting low-temperature phases has long been an active topic of research~\cite{ma:lee, finkelstein, inhomogeneities, fisherbh, haviland, me:nadya, recent}: although a quantum superconductor-insulator transition (SIT) is generally believed to occur, the actuating mechanism(s), the nature of the insulating phases and of their charge carriers, and the possible existence of a zero-temperature metallic regime~\cite{finkelstein, me:nadya} are still open and much-discussed questions~\cite{recent, feigelman, mezard}. Theories of the transition vary, e.g., in whether they incorporate the composite nature of Cooper pairs and the repulsive and/or long-range parts of the Coulomb interaction, and in the nature of the disorder considered; the importance of each of these factors in the condensed-matter systems studied is not, in general, understood \textit{a priori}. 

It is thus hoped that the microscopically well-characterized settings provided by ultracold atomic gases, trapped in controllably disordered environments~\cite{lewenstein, demarco}, can shed light on the SIT. Beyond this general motivation for studying the SIT in ultracold atomic gases, we note that two major approaches to the SIT correspond to the limits of BCS and BEC pairing respectively---namely, (i)~those that investigate the BCS pairing of localized electrons, ignoring phase fluctuations and repulsion~\cite{ma:lee, feigelman}; and (ii)~those that address tightly bound, repulsively interacting bosons via (e.g.) the Bose-Hubbard model, ignoring their composite character~\cite{fisherbh}. As these approaches correspond respectively to the BCS and BEC limits, interpolating between them is equivalent to understanding how the SIT evolves as one tunes across the BEC-BCS crossover via a Feshbach resonance~\cite{feshbach:review}. This question has not, to our knowledge, been previously considered in detail; the limiting cases are discussed in Ref.~\cite{shklovskii}, and disorder effects on the superconducting transition temperature have been treated perturbatively~\cite{orso}. As these works observe, superconductivity is more robust to disorder in the BCS regime than in the BEC regime, owing to Anderson's theorem~\cite{anderson}; however, the properties of the SIT in the intermediate, near-unitary regime have received little attention. 

In the present work we address the nature of the SIT in the intermediate region, focusing on the morphology of the zero-temperature phase diagram. Our treatment is quantitatively valid for narrow Feshbach resonances and/or dense gases~\cite{gurarie, ho:narrowres}; the former of these limits can be realized in experiments with $^6$Li~\cite{esslinger12} (which has a narrow resonance at approximately 543 G~\cite{gurarie, hulet}). In the narrow-resonance limit, we show that the superconductor-insulator phase boundary is, for a certain range of disorder strengths, nonmonotonic as a function of detuning from the Feshbach resonance, involving an intermediate insulating phase for detunings slightly below the mobility edge. We establish this central, unexpected, result both for bounded disorder and for unbounded Gaussian disorder (which is most relevant, e.g., to optical speckle potentials). We argue that the shape of the phase diagram can be understood in terms of a competition between the pairing instability and the tendency of Cooper pairs, which are bosons, to condense into deeply localized states; thus, the intermediate insulating phase has fermions of large localization length coexisting with deeply localized bosons. We discuss the feasibility of realizing this intermediate insulating phase, and finally note possible connections between our analysis and the $\epsilon$-expansion approach pioneered by Nishida and Son~\cite{nishida}.

\section{Model and approach}

We consider the two-channel model of a Feshbach resonance~\cite{gurarie}. This model involves fermionic gases consisting of consisting of two ``spin'' species (i.e., internal states), coupled to a gas of bosonic molecular states, such that two fermions can pair up to form a bosonic molecule:

%
\bea\label{eq:model}
H & = & \int d^d x \sum_\alpha \psi_\alpha^\dagger(\mathbf{x}) \left( - \frac{\nabla^2}{2M} + V(\mathbf{x}) - \mu \right) \psi_\alpha(\mathbf{x}) \nonumber \\
& & \quad + \phi^\dagger(\mathbf{x}) \left( - \frac{\nabla^2}{4M} + V(\mathbf{x}) + \varepsilon - 2 \mu \right) \phi(\mathbf{x}) \\
&& \quad + U (\phi^\dagger(\mathbf{x}) \phi(\mathbf{x}))^2 + g (\phi^\dagger(\mathbf{x}) \psi_\uparrow(\mathbf{x}) \psi_\downarrow(\mathbf{x}) + \mathrm{h.c.}) \nonumber
\eea
This Hamiltonian was analyzed for clean systems in Ref.~\cite{gurarie} in the small-$g$ limit; note, however, that a weakly-coupled two-channel description applies near unitarity for \textit{all} short-range fermionic interactions near four dimensions~\cite{nishida}. For disordered systems, we have supplemented the Hamiltonian of Ref.~\cite{gurarie} with a contact repulsion of strength $U$ among the bosons, which is assumed to be smaller than the other energy scales; this contact term, always physically present, can be neglected for clean systems, but must be included for disordered systems in order to circumvent the pathological behavior of the disordered noninteracting Bose gas. As we shall see, our results are insensitive to $U$. The potential $V(\mathbf{x})$ includes both disorder and (in principle) a periodic potential (note, however, that the narrow-resonance model does \textit{not} correspond to the attractive-$U$ Hubbard model, as pair hopping is explicitly included in $H$ and does not arise by a superexchange mechanism). We shall focus on three- (or higher-) dimensional systems, in which the single-particle (i.e., $g = 0$) Hamiltonian has both localized and extended eigenstates, separated by a mobility edge~\cite{kramer} at some energy $E_c$. For most of the present work, we shall take the disorder distribution to be specified through $E_c$ (which scales with the disorder strength) and the single-particle density of states $\rho(E)$. We first analyze the case of a bounded disorder distribution without spatial correlations (e.g., a box distribution~\cite{kramer}) and then turn to the case of unbounded Gaussian disorder with short-range correlations (which is relevant, e.g., to the current speckle-potential-based experimental realizations~\cite{demarco}). While our work does not explicitly treat the case of smoothly varying or correlated disorder, such an extension would be straightforward. 

\begin{figure}[t]
	\centering
		\includegraphics{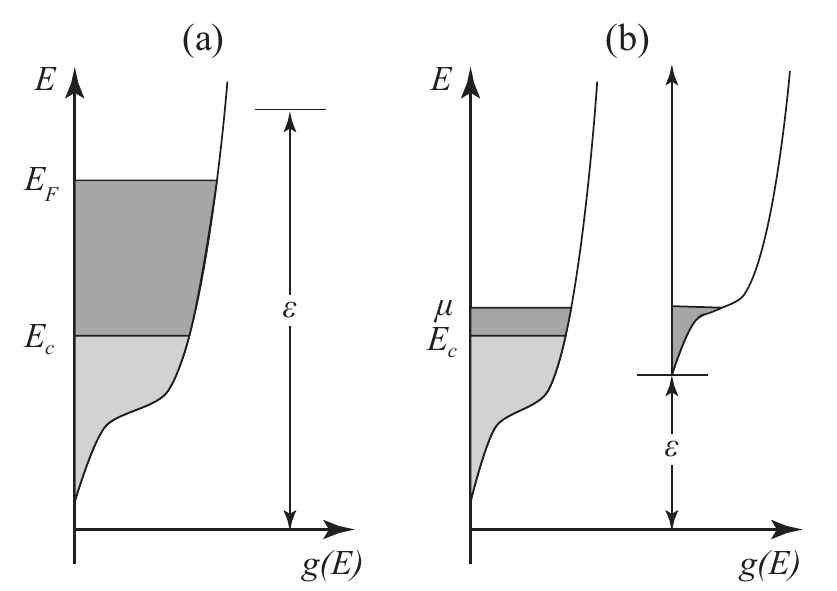}
		\caption{Limit of a narrow Feshbach resonance. (a)~Deep in the BCS limit, the bosonic channel is above $E_F$ and is therefore unoccupied; all the particles are in the Fermi channel. The Fermi energy $E_F$ is taken to be above the mobility edge $E_c$. (b)~Crossover region: $\varepsilon$ is lower than $E_F$; therefore, fermions with energies above $\varepsilon$ pair and migrate into the molecular channel. The molecules cannot all populate the lowest bosonic state owing to repulsive interactions between them; therefore, they occupy a finite energy range.}
	\label{fig:narrowres}
\end{figure}

Our approach is as follows. First, we consider the Hamiltonian $H$ in the limit $g = 0$ (known as the infinitely-narrow resonance limit~\cite{gurarie}), in which it consists of decoupled fermionic and bosonic channels (Fig.~\ref{fig:narrowres}). The equation of state for each of these channels is known, and these results can be combined to form a ``zeroth-order'' theory valid throughout the crossover. Having established the structure of this $g = 0$ theory, we include the effects of $g$ perturbatively, to lowest order, adapting the results of Ref.~\cite{feigelman}. This perturbative analysis shows the existence of regimes of reentrant superconductivity, which, we argue, offer a conservative estimate of the parameter range in which reentrant behavior takes place. The primary motivation for choosing the narrow resonance approach is that it enables us to study the entire BEC-BCS crossover while working at strong disorder (in contrast with previous works~\cite{orso}).

\section{Limiting Cases}

In this section we briefly outline the general properties that the SIT is expected to exhibit in the BEC and BCS limits, without specific reference to the narrow-resonance model. For the purposes of this section we shall assume that there is no underlying lattice. We then proceed, in the following sections, to show that the narrow-resonance approach both reproduces these expectations and provides a useful scheme for interpolating between them. 

\textit{BCS limit}. In the BCS limit, the nature of the ground state depends on the location of the Fermi energy $E_F$ with respect to the mobility edge $E_c$~\cite{ma:lee, feigelman, shklovskii}. In particular, if $E_F > E_c$ then for any infinitesimal attraction (i.e., any detuning) the ground state is superconducting~\cite{ma:lee}. On the other hand, if $E_F < E_c$ then a finite minimum attraction~\cite{feigelman} is necessary to give rise to a superconducting state. Thus, for $E_F < E_c$ an insulating ground state obtains deep in the BCS regime, but as the detuning is decreased (and the BCS coupling becomes stronger) a transition to a superconductor takes place. Thus, at fixed disorder, the critical density behaves as in the ``BCS'' side of Fig.~\ref{fig:phasediagb}(a): it is asymptotically given by the criterion $E_F = E_c$, but closer to resonance the critical density decreases.

\textit{BEC limit}. In the BEC limit, the SIT is governed by a competition between the repulsion between the Cooper pairs and the disorder strength. For weak repulsive interactions the bosons occupy isolated deep wells of the disorder potential; as the repulsive interaction strength (or density) is increased, they occupy more and more wells of the disorder potential, until eventually the Josephson coupling between the wells is strong enough to give rise to global phase coherence. An approximate criterion~\cite{shklovskii, nattermann:prb} for the critical repulsive interaction strength is given by $E_c \approx n a_b/(4M)$, where $a_b$ is the boson-boson scattering length (which decreases to zero as one tunes away from resonance). Thus, sufficiently deep in the BEC limit, the ground state is always insulating, as shown in the ``BEC'' side of Fig.~\ref{fig:phasediagb}(a); this is in sharp contrast with the BCS limit discussed above.

\textit{Crossover}. The considerations above suggest that, for a given disorder strength, the lowest critical density for the SIT occurs in the crossover regime (or, equivalently, that superconductivity is most robust against disorder in the crossover regime). Thus, the critical disorder is evidently a non-monotonic function of density, with quite different functional forms in the two limits. In what follows, we shall discuss this nonmonotonic function at a more quantitative level.

\section{Case of Bounded Disorder}

We first address the conceptually simple case of bounded disorder, and then extend our results to uncorrelated Gaussian disorder. As discussed above, we consider the decoupled ($g = 0$) limit first, then work perturbatively in $g$.

\subsection{Infinitely narrow resonance limit}

We begin with the limit of an infinitely narrow resonance (i.e., $g \rightarrow 0$), so that the fermions and bosons are decoupled. (Following Ref.~\cite{gurarie}, we assume that the $g \rightarrow 0$ limit is taken in such a way that the two channels equilibrate with each other.) For $\varepsilon \gg E_F$ (i.e., in the BCS limit), the Fermi sea is filled up to a Fermi energy $E_F$; depending on whether this exceeds $E_c$, the system is either an insulator or a metal. On the other hand, for $-\varepsilon \gg E_F$ (i.e., in the extreme BEC limit), all fermions are tightly bound into molecules interacting via the weak contact repulsion $U$. The case of $U = 0$ is pathological, as it involves all bosons occupying the lowest single-particle state; for small nonzero $U$, however, the bosons form a ``Bose glass'' consisting of widely separated, mutually phase-incoherent, puddles of bosons at deep disorder minima: thus, the system is insulating regardless of the relative locations of $E_F$ and $E_c$. We shall assume that $U$ (or the corresponding scattering length $a_0$) is small enough, and the relevant densities low enough (i.e., $ n a_0 / 2m \ll E_c$), that the molecules are located in the exponentially rare Lifshitz tail states~\cite{kramer} of the disorder potential. 

\begin{figure}[t]
	\centering
		\includegraphics[height=2.2in, width=3.3in]{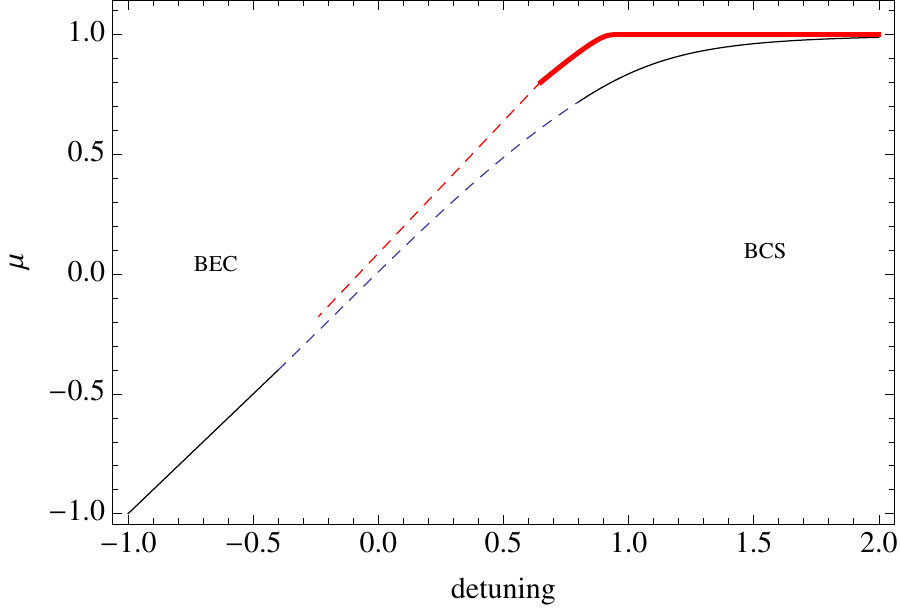}
		\caption{Change of the chemical potential $\mu$ with detuning, in the infinitely narrow resonance limit for finite $U$, in the case of bounded disorder (red dashed line) and unbounded Gaussian disorder (black solid line). Energies are expressed relative to the Fermi energy. The two cases are similar in the BEC regime but differ in the BCS regime.}
	\label{fig:mumoves}
\end{figure}

In the $g \rightarrow 0$ limit, the crossover between these limits occurs as follows. As one decreases $\varepsilon$ until it is lower than $E_F$, the chemical potential $\mu$ begins to track $\varepsilon$, as it is favorable for fermions with energies greater than $\varepsilon$ to pair up into molecules, which then form a Bose glass. In contrast with the clean case~\cite{gurarie}, we must self-consistently determine the chemical potential, as the bosons trapped in deep rare wells interact. We first determine the chemical potential due to a dilute gas of interacting bosons at density $n_b$, adapting the analysis of Ref.~\cite{nattermann:prb} as detailed in the Appendix; the result is:

\beq\label{chempot}
\mu \approx \varepsilon + \mathcal{E} \left[ \ln \frac{n_0}{n_b}\right]^{-2/3}
\eeq
where $\mathcal{E}$ is the characteristic energy scale associated with the disorder, and $n_0 \sim a_b^{-1/3}$, where $a_b$ is the scattering length for bosons. (Note that the chemical potential therefore depends weakly on $U$.) Given this expression, one can determine the density of bosons in the intermediate regime, assuming they are able to equilibrate with the fermions, by setting the two chemical potentials to be equal, and noting that the number of molecules created is one half the number of fermions depleted from the Fermi sea:

\beq\label{nb}
n_b(\mu) = \frac{1}{2} \int_\mu^{E_F} \rho(E) dE,
\eeq
where $\rho(E)$ is the density of states for the fermions. Note that Eqs.~(\ref{chempot}, \ref{nb}) together comprise an equation of state for the composite system in the $g = 0$ limit; the dependence of the chemical potential on detuning is shown in Fig.~\ref{fig:mumoves}. The qualitative behavior they imply is as follows. In the ``metallic'' case where $E_F > E_c$, $\mu$ decreases past $E_c$ for some $\varepsilon$; at this point, the fermionic sector undergoes a metal-insulator transition, and consequently (as the bosons are trapped in isolated wells) the composite system becomes insulating. On the other hand, in the ``insulating'' case where $E_F < E_c$, the system is always insulating. Therefore, in this limit one is led to the schematic phase diagram denoted by the black dashed lines in Fig.~\ref{fig:phasediagb}(a).

The above discussion extends trivially to the case of a lattice with one or more bands; each band generically has an upper and a lower mobility edge, and metal-insulator transitions occur whenever the chemical potential (which is approximately $\varepsilon$) coincides with a mobility edge. For sufficiently small $U$, bosons always occupy localized states near the bottom of the lowest band, and do not contribute to transport.

\subsection{Finite-width resonances: BCS and BEC limits}

We now consider how the above picture is altered for small but finite $g$. We first briefly sketch how the narrow-resonance formalism reproduces the known qualitative features of both the BCS and BEC regimes~\cite{shklovskii}, and then turn to the intermediate regime. 

a. \textit{BCS regime, $\varepsilon \gg E_F$}. In this regime, there is a gap to the creation of molecules; therefore, the bosonic channel can be integrated out. The nature of the ground state depends on the position of $E_F$ relative to $E_c$. For $E_c < E_F$, the Fermi energy is among the extended states (i.e., the $g = 0$ state is metallic), and BCS pairing occurs for arbitrary $g > 0$ and arbitrary $\varepsilon$, with a gap scale set by $\Delta \simeq E_F \exp[-\varepsilon/(\rho(E_F) g^2)]$. For $E_c > E_F$, the Fermi energy is among the localized states; pairing occurs in this regime, but only for couplings above some critical value $g_c$ (or, more relevantly to the present discussion, above some critical value $\varepsilon_c$), for which a rough estimate is given by the following implicit expression~\cite{feigelman, mezard}:

\beq\label{feigel}
\left( \frac{E_c - E_F}{E_c} \right)^{3\nu} \frac{g_c^2}{E_F \varepsilon_c} \exp[\varepsilon_c/(\rho(E_F) g_c^2)] \approx 1.
\eeq
where $\nu$ is the localization length critical exponent, which is approximately unity~\cite{kramer}. Thus, at zero temperature the system is insulating in the deep BCS regime ($\varepsilon \rightarrow +\infty$), but undergoes an SIT upon sweeping toward resonance, as $\varepsilon$ decreases and the effective superconducting coupling $g^2 / \varepsilon$ correspondingly increases. 

b. \textit{BEC regime, $\varepsilon \ll 0$}. In this regime, all fermions are dimerized and there is a gap to the breaking of molecules; therefore, the fermions can be integrated out, giving rise, for any $g > 0$, to an effective repulsive interaction among the molecules. The strength of this induced interaction depends on the associated energy denominator. In particular, deep in the BEC regime, the induced interaction is parameterized by the scattering length~\cite{gurarie}

\beq\label{eq:ab}
a_b = \frac{g^4 m^{5/2}}{32 \pi^2 |\varepsilon|^{3/2}},
\eeq
which decreases as $\varepsilon \rightarrow -\infty$. Note that, while the narrow-resonance behavior of $a_b$ differs quantitatively from the broad-resonance value~\cite{gurarie}, the \textit{trend} that interactions weaken far from the resonance is true in both cases. Moreover, as previously discussed, our results for the disordered system depend only logarithmically on $a_b$, so that the distinction between the narrow- and broad-resonance scattering lengths is immaterial for our  purposes. Note that if the quantity 

\beq
\frac{ n_b(\mu) a_b}{4 m} 
\eeq
exceeds the characteristic disorder scale $\mathcal{E}$, the Bose glass undergoes a transition to a superfluid state. Consequently, the ground state in the BEC regime is a superfluid for sufficiently high densities \textit{and} interaction strengths, whereas in the BCS regime it is a superfluid at high densities, \textit{regardless} of the interaction strength. 

\subsection{Intermediate regime, $0 < \varepsilon < E_F$}

The intermediate regime poses greater difficulties: in this regime, the system possesses both fermionic and bosonic excitations at the chemical potential, so that, in principle, neither channel can safely be integrated out. The argument of the present section is that one can nevertheless compute certain properties of the phase diagram in this regime, and show, in particular, that the critical coupling is a nonmonotonic function of $\varepsilon$. For specificity, we shall focus during this argument on the case of a three-dimensional Anderson lattice model~\cite{kramer}, with on-site energies drawn from the uniform distribution $[-W t/2, W t/2]$, where $t$ is the nearest-neighbor fermionic hopping matrix element; however, the results are evidently more generally applicable. (The bosonic hopping is $\sim t/2$ for this model; note, again, that the two-channel model on a lattice is distinct from the attractive-$U$ Hubbard model.)

We assume that $W$ is close to but not quite at its critical value~\cite{cuevas}, $W_c \approx 16.5$, so that $E_c$ is near the center of the band; we assume, further, that $E_F$ is slightly above $E_c$. The density of states $\rho(E)$ is known~\cite{edwards:thouless} to be analytic around $E_c$, so it follows that, if $E_c$ is near the center of the band, $\rho(E)$ is approximately constant near $E_c$. Under these conditions, the ground state is superfluid for infinitesimal $g$ in the BCS regime, as previously discussed; however, as $\varepsilon$ is decreased until $\mu < E_c$, the ground state becomes insulating (in both the fermionic and bosonic sectors) in the decoupled limit, $g \rightarrow 0$. For $g \rightarrow 0$, there are two kinds of low-energy states: (a)~fractal, almost-extended fermionic states, and (b)~deeply localized Lifshitz tail~\cite{kramer, johri:bhatt} states, each occupying a rare fluctuation of the disorder potential and holding a small number of bosons (these become arbitrarily rare as $E_F \rightarrow E_c$). As both channels are stable against superfluidity for infinitesimal $g$, a minimum inter-channel coupling $g_c > 0$ is required for global superfluidity to occur; and under the assumptions above it is clear that, at $T = 0$, global superfluidity first arises via a fermionic pairing mechanism as $g$ is increased through $g_c$. We argue that, under the assumptions above, $g_c$ is in general a non-monotonic function of $\mu$ and therefore of $\varepsilon$.

We now discuss how pairing arises in this regime, and estimate the pairing strength. In contrast to the BCS and BEC limits, the intermediate regime requires us to consider \textit{which} bosonic states mediate the pairing interactions---in the clean system, one can assume that pairing is mediated by the zero-momentum bosonic state~\cite{gurarie}, but this has no direct analog in the disordered system. To address this issue, we first note that there are, in principle, two kinds of bosonic states that can mediate pairing: (i)~the exponentially rare, but strongly interacting Lifshitz tail states within $g$ of the chemical potential, and (ii)~the more abundant but relatively weakly interacting states with appreciable gaps. We argue that, under the assumptions listed above, the Lifshitz-tail contribution to pairing is negligible, so that all the \textit{relevant} bosonic degrees of freedom are in fact gapped and can be integrated out, allowing us to adapt the BCS-limit results with some effective energy denominator $E_{LT}$. (We shall qualitatively address the role of the tail states in the next subsection, and show that they \textit{accentuate} the nonmonotonicity of the phase diagram.) 

The argument for neglecting the Lifshitz tail states runs as follows. The resonant pairing mediated by the tail states within a window $dE$ of the chemical potential can be estimated as 

\beq
M_{\mathrm{res.}} \approx g \sqrt{n_b} P_b \times \mathrm{overlap}
\eeq
where $n_b$ is the density of bosons; $P_b$ is the number of resonant tail states $\sim \exp[- B (\mu - \varepsilon))^{-3/2}]$, where $B$ is related to $\mathcal{E}^{3/2}$;  and the coupling is modulated by a typical overlap integral (of the form $\int |\psi|^2 \phi$) associated with the fermion-boson coupling [Eq.~(\ref{eq:model})]. By contrast, the off-resonant pairing mediated by higher-energy bosonic states within an energy window $dE$ can roughly be estimated as 

\beq
M(E) \sim \frac{g^2}{E - \mu} \rho_b(E) dE \times \mathrm{(overlap)}^2,
\eeq
where $\rho_b$ is the bosonic density of states at $E$. Comparing these two expressions, one finds that (even ignoring the nature of the overlaps), the exponential increase in the number of states outweighs the decreased coupling to each state, so long as $g / (Wt) \gg \sqrt{N_b} P_b(\mu)$, which is always true for $\mu \approx E_F$ (i.e., $N_b \rightarrow 0$). Furthermore, the numerical results of Ref.~\cite{cuevas} suggest that wavefunctions far apart in energy are in fact spatially \textit{anticorrelated}. This effect would further suppress the interactions between the fermions and the tail-state bosons, and therefore extend the regime of quantitative validity of the present analysis.

Having established that the tail states contribute negligibly to pairing, we now estimate which states do provide the dominant pairing interactions. One can easily see that states that are too high in energy cannot contribute to pairing, as for these states the matrix elements of the form $\int dx |\psi(x)|^2 \phi(x)$ are products of three rapidly oscillating functions and must vanish. Combining these observations (viz. that both states with very small energy denominators and those with large energy denominators are negligible), one sees that $M(E)$ must be peaked at some $E_{LT}$ measured from $\varepsilon$. This energy $E_{LT}$ is of order $W t$ in general; for the case of the three-dimensional Anderson model, the considerations of Ref.~\cite{cuevas} (regarding the ``stratification'' of coordinate space) would suggest a value $E_{LT} \approx \varepsilon + W t / 2$~\footnote{The argument can be summarized as follows. Ref.~\cite{cuevas} argues that states within $\sim t$ of the mobility edge have appreciable spatial overlap with the fractal states, in the sense of having ``support'' on the same lattice sites. The fermions would primarily interact with the lowest, nodeless bosonic modes supported on these sites.}. However, for our purposes we only need the fact that it is nonzero, which enables us to estimate a BCS pairing interaction of strength $\gamma(\mu) \sim g^2 / (E_{LT} - \mu)$. It is crucial to note that $\gamma(\mu)$ \textit{increases} as one tunes toward resonance, because $E_{LT} = \varepsilon +$~constant, and $\mu - \varepsilon$ increases as the tail states fill up with bosons. 

The arguments above lead us to a problem involving only fermionic states at the chemical potential and virtual bosonic states $\sim E_{LT}$ away from it. This reduced problem can now be addressed exactly as in the BCS limit, by inserting $\gamma(\mu)$ into Eq.~(\ref{feigel}), to arrive at the following criterion for superfluidity:

\beq\label{feigel2}
\left( \frac{E_c - \mu}{E_c} \right)^{3\nu} \rho(\mu)\gamma(\mu) \exp[1/(\rho(\mu)\gamma)] \alt 1.
\eeq
In this expression, the relation between $\mu$ and $\varepsilon$ is given by Eqs.~(\ref{chempot}, \ref{nb}). Thus, Eq.~(\ref{feigel2}) thus determines a critical value of $\gamma$ and thus of $g$, which is plotted in Fig.~\ref{fig:phasediagb}(d), and can be seen to be nonmonotonic, exhibiting a transition to an insulator very close to $E_c$ as well as a reentrant superfluid at lower detunings [Fig.~\ref{fig:phasediagb}(c)-(e)]. The intuitive reason for this non-monotonic behavior is that the bosonic tail states fill up rapidly (there being very few such states), and hence decrease the energy denominator for pairing ($E_{LT} - \mu$) without substantially depleting the Fermi sea. Note that as $\varepsilon$ is swept closer still to resonance, the fermionic density of states at $\mu$ rapidly decreases, triggering a second transition to an insulating state, which persists into the BEC limit. This second transition generally occurs far from the mobility edge, and is thus beyond the scope of the present work. (However, it follows from our considerations and the previously mentioned results for the BEC limit~\cite{shklovskii} that such a transition must occur at some point.)

Let us briefly review the argument given above for the shape of the phase diagram. Consider a situation in which $E_F$ is slightly above $E_c$, and the detuning $\varepsilon$ is tuned so that $\mu$ is slightly below $E_c$. The ground state is an insulator for $g = 0$, and this insulator is evidently stable for sufficiently small, nonzero $g$; however, it becomes a superconductor at some $g_c$. This critical coupling $g_c$ depends on $\mu$ (or equivalently on $\varepsilon$) in three ways: (1)~the localization length at $\mu$ decreases as $\mu$ decreases, suppressing pairing; (2)~the fermionic density of states $\rho(\mu)$ decreases as $\mu$ decreases, suppressing pairing; and (3)~the pairing interaction grows stronger as $\mu$ decreases, owing to the concomitant decrease in the energy denominator $E_T - \mu$. Sufficiently close to the mobility edge, effect~(1) dominates; sufficiently deep in the insulating phase, effect~(2) dominates; in either case pairing is suppressed with decreasing $\mu$. However, as argued in this work, there is a range of \textit{intermediate} energies for which effect~(3) is the dominant effect, so that pairing is \textit{enhanced} by a decrease of $\mu$, leading to a nonmonotonic phase diagram overall. (The competition between effects (2) and (3) has previously been noted~\cite{cohen}, in the condensed-matter context, as leading to a maximum in the transition temperature of SrTiO$_3$ as a function of carrier density.)

\subsection{Role of the Lifshitz tail states in mediating pairing}

In the preceding discussion, we neglected the role played by the Lifshitz tail states in mediating pairing, arguing that this role was negligible for $E_F \approx E_c$. We now return to this point. The role of the Lifshitz tail states is to provide an additional channel for pairing, of strength $\sim \sqrt{n_b}$ where $n_b$ is the density of molecules. Now, as $\varepsilon$ is decreased, the bosonic band fills up, so that both $\rho_b$ and $n_b$ increase. Thus, the coupling strength increases \textit{faster}, as $\varepsilon$ (or equivalently $\mu$) is reduced, than our previous analysis predicted. This strengthens effect~(3) discussed in the previous paragraph, and thus accentuates the nonmonotonicity of the phase diagram. 

\subsection{Phase diagram}

\begin{figure}[ht]
	\centering
		\includegraphics{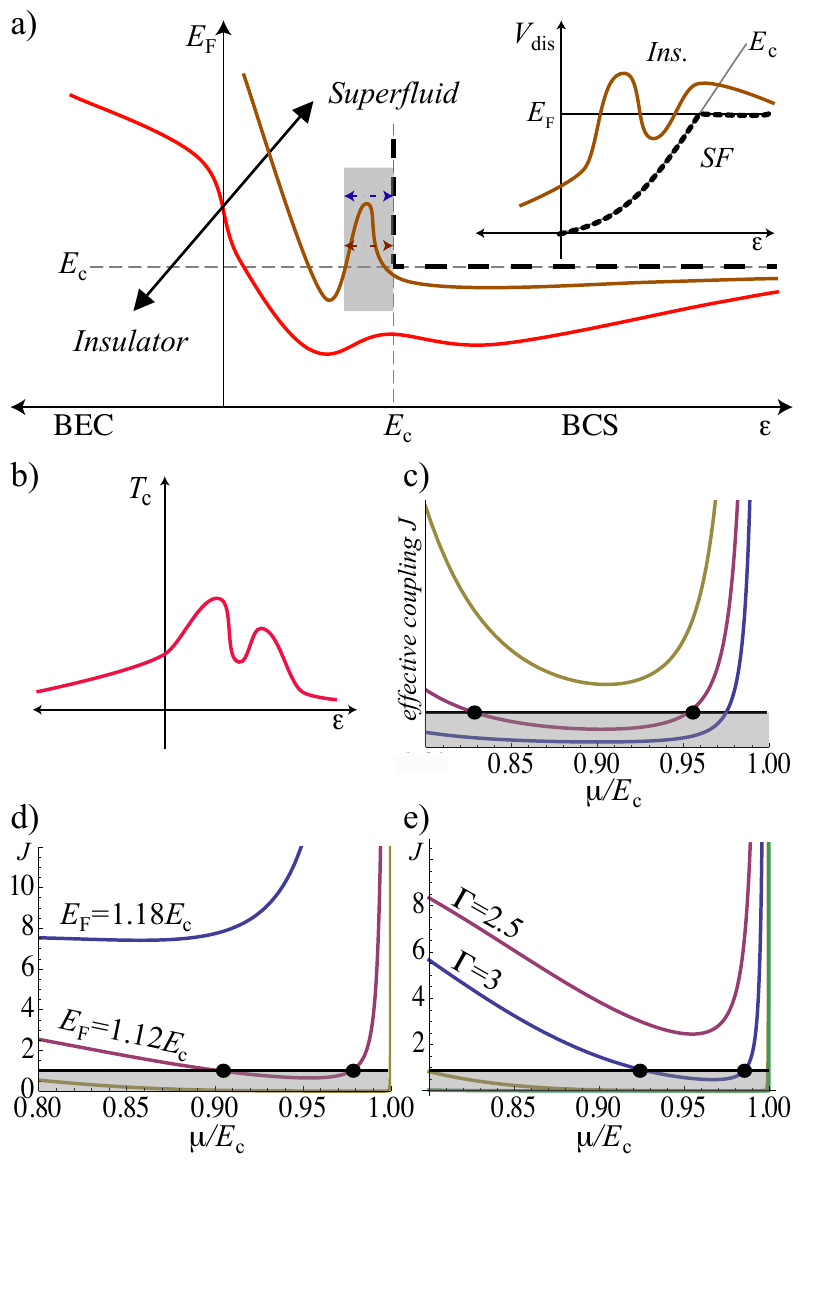}
	\caption{(a)~Schematic $T=0$ phase diagram of the SIT as a function of detuning. Main panel shows the phase diagram at fixed disorder strength as density (i.e., $E_F$) is varied: black dashed line corresponds to an infinitely narrow resonance, whereas the brown and red lines correspond to relatively narrow and broad resonances respectively. $E_c$ is the mobility edge. The results of this paper concern the shaded region. Inset shows phase diagram as  a function of disorder at fixed density; color coding is the same as in the main panel. (b)~Sketch of $T_c$ as a function of detuning near an SIT; the proximity to an insulating phase is expected to depress $T_c$. (c)-(e)~Inverse of the left-hand side of Eq.~(\ref{feigel2}) as a function of $\mu$. Panel (c) shows the case of bounded disorder at three densities, from $E_F = 1.18 E_c$ (top) to $E_F = 1.1 E_c$ (bottom), at fixed $\Gamma \equiv \rho(E_c) g^2 / E_c = 1/20$; panel (d) shows the case of Gaussian disorder, with $\rho(E_c) g^2 / E_c = 1/3$; and panel (e) shows the case of Gaussian disorder, with $E_F = 1.1 E_c$ and different values of $\Gamma$. The insulating region is shaded gray.}
	\label{fig:phasediagb}
\end{figure}

The considerations in the previous sections can be synthesized into a global phase diagram, at fixed disorder strength, in terms of the density, detuning, and coupling $g$ (i.e., the resonance width). The results are sketched in Fig.~\ref{fig:phasediagb}(a)-(c). In the narrow-resonance limit, there are four possible scenarios for the BEC-BCS crossover, as follows. (1)~When $E_F \ll E_c$, the system stays insulating through the crossover. (2)~When $E_F$ is near, but lower than $E_c$, superfluidity occurs at some finite, positive detuning $\varepsilon$, before vanishing at yet smaller $\varepsilon$. (3)~When $E_F$ is slightly higher than $E_c$, superfluidity is destroyed for $\varepsilon$ just below the mobility edge, but reappears as the detuning is decreased further, before finally vanishing again in the BEC limit. (4)~For $E_F \gg E_c$, the ground state is superfluid in the BCS and intermediate regimes, and superfluidity persists well into the BEC regime before eventually disappearing deep in the BEC regime, as the interactions between bosons weaken. Of these scenarios, the most exotic is Scenario~(3), and one might ask to what extent it is limited to the narrow-resonance approach adopted here. The crucial point in our arguments above where the narrow-resonance limit was invoked was in justifying non-degenerate perturbation theory for the pairing interaction in the intermediate regime; while strictly valid only for $g \ll E_{LT} \sim E_c$, one expects that it will remain reasonably accurate so long as $g \alt E_c$. (By contrast, in the more commonly studied broad-resonance limit, $g \gg E_F \approx E_c$, so that the variation of the energy denominator is irrelevant and one does not expect a re-entrant superconducting phase.) However, this limitation can be circumvented either by working at high densities and strong disorder, or by choosing a narrow Feshbach resonance (as discussed below); besides, even beyond its regime of quantitative validity, the underlying mechanism by which strong pairing is offset by the tendency of bosons to localize might lead, e.g., to anomalous behavior in the critical temperature of a disordered Fermi gas near unitarity.   

\subsection{Nature of insulator}
 
The discussion above suggests a simple physical picture of the insulating phase in the intermediate regime: viz. that it consists of two interpenetrating insulators, a Lifshitz glass~\cite{nattermann:prb, lewenstein} and a ``fractal insulator''~\cite{feigelman}, which are essentially \textit{mutually noninteracting} because of the lack of wavefunction overlap. Thus, the gap and other properties of the insulator---measured, e.g., via rf spectroscopy~\cite{rf}---near the SIT should consist of an appropriately weighted average of the properties of these two systems. The coupling between the systems grows for smaller $\varepsilon$ and/or higher densities, as the chemical potential then no longer lies in the bosonic Lifshitz tail (defined as the region in which the density of states grows exponentially with energy); in this regime, a more suitable model is to treat the fermions as being strongly proximity-coupled to puddles of bosonic superfluid~\cite{me:nadya}.

\section{Unbounded Gaussian disorder}

We now extend our results to the case of uncorrelated Gaussian disorder of the same characteristic energy scale $\mathcal{E}$, the case most directly relevant to recent experiments in optical lattices~\cite{demarco}. The essential distinction between bounded and unbounded disorder is the nature of the Lifshitz tail~\cite{kramer}; in the uncorrelated Gaussian case, the density of states in the tail scales as $\rho(E) \sim \exp(-B |E|^{1/2})$. For unbounded disorder, there is some finite probability of finding states at arbitrarily low energies, so the bosonic band is not bounded below. Thus, the BCS and BEC regimes are not sharply separated: at any detuning, there are \textit{some} molecular states as well as \textit{some} fermionic states, and the system is always in the intermediate regime discussed above. To extend the previous analysis to this case, it suffices to replace Eq.~(\ref{chempot}) with the following expression~\cite{nattermann:prb}:

\beq\label{chempot2}
\mu = \varepsilon - \mathcal{E} \left[\ln\left( \frac{E_F}{ E_F - \mu}\right)\right]^2
\eeq
Note that Eq.~(\ref{chempot2}), unlike Eq.~(\ref{chempot}), predicts that, regardless of $\varepsilon$, $\mu < E_F$ (although the difference vanishes as $\mathcal{E} \rightarrow 0$ in the clean system limit). This replacement changes various quantitative features of the phase diagram, but our qualitative conclusions (and in particular, the possibility of a nonmonotonic superfluid-insulator boundary) still hold (see Fig.~\ref{fig:phasediagb}(d)-(e)).

Another frequently considered, unbounded disorder distribution is the Lorentzian distribution, for which the density of states can be computed exactly~\cite{lloyd}; however, this distribution does not possess an exponential Lifshitz tail (instead, the tail density of states falls off as $\rho(E) \sim 1/E^2$). Thus, the relative contribution of tail states to pairing is much larger than in the bounded or Gaussian cases. While the qualitative considerations of the previous section are expected to apply here, the parameter-dependence of the transition line is expected to differ substantially.  


\section{Discussion}

This work has addressed the behavior of the superfluid-insulator transition for disordered Fermi gases across a BEC-BCS crossover, focusing on the intermediate (i.e., near-unitary) regime. We have established that the superfluid-insulator phase boundary is nonmonotonic as a function of detuning, so that an experiment sweeping across a Feshbach resonance can in principle see three separate transitions. We were able to establish these conclusions quantitatively only in the narrow-resonance regime (i.e., where the energy scale associated with the Feshbach resonance width~\cite{gurarie} is smaller than the Fermi energy); however, it is plausible that, as long as the resonance is not too broad (i.e., if $g \approx E_F$), the proximity to an insulating phase will manifest itself in a nonmonotonic critical temperature. Moreover, narrow Feshbach resonances do exist (e.g., $g/E_F \approx 0.1$ in $^6$Li~\cite{gurarie, hulet}), and have been proposed as a way of realizing higher-temperature superconducting transitions in ultracold atomic systems~\cite{ho:narrowres}. Finally, one should note that, as the width of the resonance is measured relative to the Fermi energy, all experiments at sufficiently high densities (e.g., in tightly confining traps) can be described by a narrow-resonance model. In order to stress the importance of the mobility edge, we have restricted our analysis to three-dimensional disordered systems; however, the generalization to systems of reduced dimensionality is straightforward. 

A separate limitation of our analysis, which applies generally to the dynamics of disordered quantum systems, is that in considering the equilibrium state we have neglected the experimentally relevant question of whether equilibrium is achieved on the relevant timescales (which must, at a minimum, be much larger than $1/g$ in the narrow-resonance limit). Indeed, recent work on the many-body localization problem~\cite{mbl} suggests that closely related systems do not equilibrate in the thermodynamic limit.

\textit{Near-unitary system in $4 - \epsilon$ dimensions}. Finally, we note that the weakly-coupled two-channel model [Eq.~(\ref{eq:model})] becomes valid for any Feshbach resonance, near unitarity, as the dimensionality of space approaches four. (The resemblance arises because the two-body wavefunction at resonance is sharply peaked when the particle positions coincide, so that weakly bound or even resonant fermion pairs behave like pointlike bosons~\cite{nishida}.) It follows that the ground state at unitarity is insulating for arbitrarily weak disorder near four dimensions. It is straightforward to see that the ground state near unitarity is also generically insulating near \textit{two} dimensions, by the following logic. Near two dimensions, unitarity corresponds to vanishingly attractive interactions; however, a finite disorder strength would localize all single particle states in two dimensions, thus overcoming an infinitesimal pairing term. As a consequence, the phase diagram and transition temperature of a disordered unitary Fermi gas are nonmonotonic as a function of dimensionality, in contrast with the clean-system limit~\cite{nishida}. That the SIT occurs for weak disorder in both the $4-\epsilon$- and $2 + \epsilon$-dimensional limits suggests the $\epsilon$-expansion as a promising strategy for future studies of the unitary-gas SIT using renormalization-group methods~\cite{nikolic}. 

\vspace{10pt}

\section*{Acknowledgments}

I am indebted to Markus M\"{u}ller, Philipp Strack, Arijeet Pal, Johannes Bauer, and Paul Goldbart for helpful discussions, and to Markus M\"{u}ller, Arijeet Pal, and Philipp Strack for a critical reading of this manuscript. This work was supported in part by the DOE Division of Materials Science under Grant No. DE-FG02-07ER46453, and in part by the Harvard Quantum Optics Center.

\appendix*

\section{Chemical potential in the bosonic channel}

We briefly sketch how the density of bosons confined in Lifshitz tail states can be estimated, for the case of bounded disorder. (The unbounded case is addressed in Ref.~\cite{nattermann:prb}.) For concreteness we assume the on-site disorder takes on two values, $\pm E_0$. Suppose the chemical potential is in the Lifshitz tail; then, according to a standard argument~\cite{kramer}, the single-particle states that can be filled must have dimensions of at least $R_- = \sqrt{2m(\mu - E_0)}$. In the Thomas-Fermi approximation (valid for the relatively large wells in the Lifshitz tail), a well of size $R < R_-$ typically holds $N_R \approx (\mu - E_0) R^3 / g$ particles; put differently, a typical well at energy $E$ holds $N_E \sim (\mu - E_0)  (E - E_0)^{-3/2}$ particles. The total number of particles is given by 

\beq
N = \int N(E) \rho(E) dE,
\eeq
where we can use the well-known result~\cite{kramer} for the density of states, $\rho(E) \sim \exp(-B |E - E_0|^{-3/2})$. This integral is dominated by the contribution from energies around $\mu$. Therefore, to logarithmic accuracy, one can invert the relation between $N$ and $\mu$ to write

\beq
\mu = E_0 + \mathrm{const.} [\ln n_0/n]^{-2/3}
\eeq
for some constant, which, on dimensional grounds, must be the energy scale $\mathcal{E}$ associated with the disorder strength.

\end{document}